\documentclass[aps,prl,twocolumn,showpacs,superscriptaddress]{revtex4}
\usepackage{amsfonts}

\usepackage{graphicx}
\usepackage{subfigure}
\usepackage{amsmath}
\usepackage{bm}

\begin{document}

% braket.sty          Macros for Dirac bra-ket <|> notation and sets {|}
% Donald Arseneau     asnd@triumf.ca     Last modified 05-Dec-1999.
% This is free, unencumbered, unsupported software.
%
% Commands defined are:
% \bra{ }   \ket{ }   \braket{ }   \set{ }    (small versions)
% \Bra{ }   \Ket{ }   \Braket{ }   \Set{ }    (expanding versions)
% 
% The "small versions" use fixed-size brackets independent of their
% contents, whereas the "expanding versions" make the brackets and 
% vertical lines expand to envelop their contents (internally using 
% the \left and \right commands).  You should use the vertical bar
% character "|" to input any extra vertical lines.  In \Braket these
% vertical lines will expand to match the arguments, and in \Set the
% first vertical will expand this way.  E.g.,
%   \Braket{ \phi | \frac{\partial^2}{\partial t^2} | \psi }
%   \Set{ x\in\mathbf{R} | 0<{|x|}<5 }
%
% NOT defined is "\ketbra" (for projection operators) because I prefer
% \ket{ } \bra{ }.
%
% Because each definition is so small, it makes no sense to have a 
% complicated generic version for many bracket styles.  Instead, 
% you can just copy the definitions and change \langle or \rangle,
% < and > to what you like.
%
\def\bra#1{\mathinner{\langle{#1}|}}
\def\ket#1{\mathinner{|{#1}\rangle}}
\def\braket#1{\mathinner{\langle{#1}\rangle}}
\def\Bra#1{\left<#1\right|}
\def\Ket#1{\left|#1\right>}
{\catcode`\|=\active 
  \gdef\Braket#1{\left<\mathcode`\|"8000\let|\BraVert {#1}\right>}}
\def\BraVert{\egroup\,\mid@vertical\,\bgroup}
% The \mid@vertical is \vrule with ordinary TeX but \middle| in eTeX.
% We always avoid a \mathchoice in making the inner vertical lines.  
% Note that \right>, prints the same as \right\rangle but is faster.  
%
% \def\ketbra#1#2{\ket{#1}\bra{#2}}
% \def\Ketbra#1#2{\left|{#1}\vphantom{#2}\right>\left<{#2}\vphantom{#1}\right|}

% \Set{...|...} Only the first | is treated specially.
%{\catcode`\|=\active
%  \gdef\set#1{\mathinner{\lbrace\,{\mathcode`\|"8000\let|\midvert #1}\,\rbrace}}
%  \gdef\Set#1{\left\{\:{\mathcode`\|"8000\let|\SetVert #1}\:\right\}}}
%\def\midvert{\egroup\mid\bgroup}
%\def\SetVert{\egroup\;\mid@vertical\;\bgroup}

% If the user is using e-TeX with its \middle primitive, use that for
% verticals instead of \vrule.
%
%\begingroup
% \edef\@tempa{\meaning\middle}
% \edef\@tempb{\string\middle}
%\expandafter \endgroup \ifx\@tempa\@tempb
% \def\mid@vertical{\middle|}
%\else
% \let\mid@vertical\vrule
%\fi

%TCIDATA{OutputFilter=LATEX.DLL}
%TCIDATA{LastRevised=Sun Sep 24 19:45:04 2006}
%TCIDATA{<META NAME="GraphicsSave" CONTENT="32">}
%TCIDATA{CSTFile=revtx4tci.cst}

%%\input{tcilatex}

%\begin{document}

\title{High Harmonic Generation in SF$_{6}$: Raman-excited Vibrational
Quantum Beats}
\author{Zachary~B.~Walters}
\affiliation{Department of Physics and JILA, University of
Colorado, Boulder, Colorado 80309-0440, USA}
\author{Stefano Tonzani}
\affiliation{Department of Chemistry, Northwestern University,
  Evanston, Illinois 60208-3113, USA}
\author{Chris~H.~Greene}
\affiliation{Department of Physics and JILA, University of
Colorado, Boulder, Colorado 80309-0440, USA}
\date{\today}

\begin{abstract}
In a recent experiment (N. Wagner \textit{et al.} \cite{wagner2006})
on SF$_{6}$,
a high-harmonic generating laser pulse is preceded by a pump pulse
which stimulates Raman-active modes in the molecule.  Varying the time
delay between the two pulses modulates high harmonic intensity, with
frequencies equal to the vibration frequencies of the Raman-active
modes.  We propose an explanation of this modulation as a quantum
interference between competing pathways that occur via adjacent
vibrational states of the molecule.
The
Raman and high harmonic processes act as beamsplitters, producing
vibrational quantum beats among the Raman-active vibrational modes that are
excited by the first pulse. We introduce a rigorous treatment of the
electron-ion recombination process and the effect of the ionic Coulomb field in
the electron propagation outside the molecule, improving over the widely-used
three-step model.
\end{abstract}

\maketitle

\affiliation{Northwestern University Chemistry Dept.,
2145 Sheridan Rd. Evanston,IL 60208-3113}

\affiliation{Department of Physics and JILA, University of
Colorado, Boulder, Colorado 80309-0440, USA}

High harmonic generation (HHG) is commonly understood as a 3 step
process \cite{lewenstein} in which an electron ionizes from a molecule, propagates in
a strong laser field, and then recombines with the parent ion while
emitting a photon.  Acceleration by the laser field
allows the electron to return with a large kinetic energy and emit
photons with energy much higher than those of the driving laser.

Although the HHG process is primarily electronic in character, recent
experiments have shown that vibrational degrees of freedom can play a
role.  In the experiment by Wagner \textit{et al.}, \cite{wagner2006} a 25
fs,  $2.4\times 10^{14}$ W/cm$^{2}$ HHG laser pulse was preceded by a
weaker 25 fs, $5 \times 10^{13}$W/cm$^{2}$ pulse which excited
Raman-active vibrations in the molecule.  The intensity of the
high harmonic light was found to oscillate with the interpulse delay
time at the excited molecular vibration frequencies.  The surprising result
was that the breathing mode,
overwhelmingly dominant in conventional Raman
experiments with this molecule, is no longer the strongest mode seen
in the HHG experiment.

\begin{figure}
\begin{center}
\includegraphics[width=3.375in]{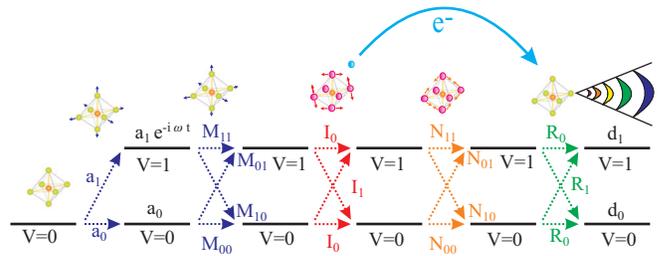}
\end{center}
\caption{(Color online) The vibrational interference model in one dimension.
The molecule ends the Raman pulse in a superposition of the v=0 and v=1
vibrational states. After a time delay, the two vibrational states are
mixed by stimulated Raman scattering, ``hopping'' during ionization
and recombination, and evolution of the ionic wavefunction while the
electron is away. Interference between adjacent vibrational states
modulates the high harmonic signal.}
\vskip 0.2in
\label{fig:crossings}
\end{figure}

In this Letter, we interpret these oscillations as an interference
between indistinguishable quantum pathways associated with different
intermediate vibrational states during the HHG process, illustrated in
Fig. \ref{fig:crossings}.  We develop a
quantum mechanical description of the recent pump-probe SF$_{6}$
experiments, using a framework that substantially improves on the three-step
model.\cite{lewenstein} We have included
the effects of the ion's Coulomb potential on the propagating
electron, and utilized a nonperturbative electron-molecule scattering
wavefunction \cite{tonzani2005} to calculate the recombination amplitude.  The resulting
calculation exhibits partial agreement with experimental observations.

We adopt a level of approximation in which all operators depending on the
nuclear coordinates are expanded to first order in the normal mode
coordinates\ $Q_{i}$; then the full vibrational state vector $\ket{\Psi
(t)} $ separates into a product of uncoupled normal mode
vectors 
$\ket{\psi(t)}^{(i)}=a_{0}^{(i)}(t)\ket{0}^{(i)}+a_{1}^{(i)}(t)\ket{1}^{(i)}$
that can be treated individually. 

After a Raman pulse with the intensity and duration used in Ref.\cite
{wagner2006}, only the $\ket{0}$ and $\ket{1}$ states of a given normal mode
have significant amplitude.  The following
two-state, one-dimensional picture shows how individual Raman-active 
modes affect the high harmonic signal. Atomic units are used throughout this
work.

During the Raman pulse, stimulated Raman scattering changes
the vibrational state of the molecule from an initial $\ket{0}$ into a
coherent superposition $a_{0}\ket{0}+a_{1}\ket{1}$ of the zeroth and
first vibrational states.  The vibrational coefficients follow
equations of motion given by
%During the Raman pulse, the molecule quickly changes from the initial state $%
%\ket{0}$ into a coherent superposition $a_{0}\ket{0}+a_{1}\ket{1}$ of the
%zeroth and first vibrational states via stimulated Raman scattering.
%Equations of motion for the vibrational coefficients are given (in a.u.) by
\begin{equation}
\begin{split}
i\dot{a}_{n_{i}}(t) =\omega _{i}\left(n_{i}+\frac{1}{2}\right)a_{n_{i}}(t)
-\frac{1}{2}\sum_{A,B}E_{A}(t)E_{B}(t) \times  \label{eq:stimulatedraman} \\
\left[ \alpha _{AB}a_{n_{i}}+\partial
_{i}\alpha _{AB}(\sqrt{n_{i}+1}a_{n_{i}+1}+\sqrt{n_{i}}a_{n_{i}-1})\right] .
\end{split}
\end{equation}
Here $\omega _{i}$ is the normal mode frequency, indices $A$ and $B$
run over $\{x,y,z\}$, $E_{A}(t)$ is the component of the electric field in
the (body-frame) $A$ direction at time $t$, $Q_{i}$ is the normalized
displacement associated with normal mode $i$ and $\alpha
_{AB}(Q_{1},Q_{2},...)$ is the polarizability tensor of the molecule. These
equations of motion have off-diagonal elements only if $\partial _{i}\alpha
_{AB}\equiv \left( 2m\omega _{i}\right) ^{-1/2}\partial \alpha
_{AB}/\partial Q_{i}|_{0}\neq 0$, which is the condition for a mode to be
Raman active. The polarizability tensor and its derivatives
 are found by performing an unrestricted Hartree-Fock calculation \cite{g98}
using the aug-cc-pVTZ basis set.
\cite{dunning1989}

Between laser pulses, $\ket{0}$ and $\ket{1}$ evolve as simple harmonic
oscillator eigenstates, becoming  
$\psi _{\text{vib}}=a_{0}\ket{0}+a_{1}e^{-i\omega \tau }\ket{1}$ at
the beginning of the high harmonic pulse for an interpulse delay of
$\tau$,
where the normal mode index $(i)$ is omitted for brevity.  
Before the electron tunnels free of
the ion, the high harmonic pulse stimulates the normal mode further
according to Eq. (\ref{eq:stimulatedraman}). This is approximated by
a unitary $2\times2$ transfer matrix $\underline{M}$, where $M_{nm}$ is the
amplitude to end in state $\ket{i}$ after starting in state $\ket{j}$ at the
beginning of the pulse. 
%Just before ionization, this gives a vibrational
%state $\left| \psi (\tau )\right\rangle =\ket{0}b_{0}+\ket{1}b_{1}$, where 
%%\begin{equation}
%$\begin{pmatrix}
%b_{0} \\ 
%b_{1}
%\end{pmatrix}
%= 
%\begin{pmatrix}
%M_{00} & M_{10} \\ 
%M_{01} & M_{11}
%\end{pmatrix}
%\begin{pmatrix}
%a_{0} \\ 
%a_{1}e^{-i\omega \tau }
%\end{pmatrix}$
%\end{equation}.

%The vibrational state evolves further when the electron ionizes from the
%molecule and when the electron recombines with the molecule. Although both
%processes primarily affect the electronic part of the molecular
%wavefunction, both are modulated strongly by displacements of the molecule
%away from its equilibrium geometry.
%Although ionization and recombination primarily affect the electronic
%part of the molecular wavefunction, both are modulated strongly by
%displacements of the molecule away from its equilibrium geometry.
%This means that the vibrational state of
%the molecule has some amplitude to change during both processes.

Ionization and recombination, both electronic processes, are both
modulated strongly by nuclear motion.
Taylor-expanding the tunnel-ionization operator $\hat{I}$ to first order
in $Q_{i}$, with the substitutions $I_{0}\hat{%
\mathbb{I}}\equiv I|_{\text{eq}}$ and $I_{1}^{(i)}\equiv \left( 2m\omega
_{i}\right) ^{-1/2}\partial \hat{I}/\partial Q_{i}|_{\text{eq}}$ and the
identity $Q_{i}=\left( 2m\omega _{i}\right) ^{-1/2}(\hat{A}_{i}+\hat{A}%
_{i}^{\dag })$, yields the first-order expansion into raising and lowering
operators, $\hat{I}=I_{0}\hat{\mathbb{I}}+\Sigma _{i}I_{1}^{(i)}(\hat{A}_{i}+%
\hat{A}_{i}^{\dag }).\;$Here $\mathbb{\hat{I}}$ is the identity
operator,  and $\hat{A}_{i}^{\dag },\hat{A}_{i}$ are the raising and
lowering operators for the $i-$th normal mode. For each mode
considered, the reduced mass $m$ is equal to the mass of a single
fluorine atom.
The recombination operator $\hat{\vec{R}}$ can be derived to first
order using identical logic.  In both cases, dependence on nuclear
positions means that the vibrational state changes along with the
electronic state.

%%rewrite JT
%While the electron is away from the ion, the ion's coupled
%vibrational-electronic wavefunction's
%evolution is governed by the Jahn-Teller Hamiltonian
%\cite{estreicher,moffitt,bersuker}.  This occurs because the three
%degenerate electronic orbitals of SF$_{6}^{+}$, which have T$_{1g}$
%symmetry are coupled to one another by the ion's vibrational
%coordinates.  The Jahn-Teller Hamiltonian, given in \cite{estreicher},
%includes off-diagonal coupling between different electronic states
%(which we find to be small).

The evolution of the ionic vibrational/electronic wavefunction between
ionization and recombination is in general quite complicated, since
the three degenerate orbitals of SF$_{6}^{+}$, which have T$_{1g}$
symmetry, are coupled by vibrational degrees of freedom
\cite{estreicher,moffitt,bersuker}.
The linear and quadratic
terms in the Jahn-Teller Hamiltonian
\cite{estreicher,moffitt,bersuker}, which governs the coupled 
vibrational/electronic evolution, are found for each $Q_{i}$ by fitting the
eigenvalues of the coupling matrix to the lowest 3 adiabatic energies of
SF$_{6}^{+}$ for different
displacements of the ion away from the maximum symmetry
configuration.  The energies are found using Gaussian's CASSCF
method and a cc-PVDZ basis set\cite{g98}.  In the notation of
\cite{estreicher}, $V_{T_{2g}}$=.001209 H/bohr,
$V_{E_{g}}$=.1420 H/bohr,
$N_{1}$=-.0362 H/bohr$^2$,
$K_{T_{2g}}$=.7288 H/bohr$^2$,
$K_{E_{g}}$=1.8486 H/bohr$^2$.  For the A$_{1g}$ mode, which does not enter
into the vibronic Hamiltonian,  an adiabatic potential
$E=V_{A_{1g}}Q_{A_{1g}}+1/2 K_{A_{1g}}Q_{A_{1g}}^2$, with
$V_{A_{1g}}$=.0645 H/bohr, 
$K_{A_{1g}}$=2.98 H/bohr$^2$ gives the potential energy surface for all
three electronic states.  An important
simplification is that the off-diagonal coupling between different
electronic states, proportional to $V_{T}$, is small and can be
neglected for the short times between ionization and recombination.

Neglecting off-diagonal coupling between electronic states, the
adiabatic potential felt by the ion in a particular electronic state
is $H_{i}=\frac{p_{i}^2}{2m} + V_{i}Q_{i} + \frac{1}{2} K_{i}Q_{i}^{2}$,
which can be rewritten to first order in the basis of oscillator states of the
neutral molecule.  Evolution of the vibrational state is given by a
transfer matrix $\underline{N}=exp(-i H(t_{\text{ret}}-t_{\text{ion}}))$.

 In a
two-state treatment, the $i-$th vibrational wavefunction of
the neutral molecule after recombination has occurred is 
$\ket{\psi_{\text{vib}}}=d_{0}\ket{0}+d_{1}\ket{1}$, where 
\begin{equation}
\begin{pmatrix}
\vec{d}_{0} & \vec{d}_{1}
\end{pmatrix}
=
\begin{pmatrix}
a_{0} & a_{1} e^{-i \omega \tau}
\end{pmatrix}
\underline{M}^{T}\underline{I}^{T}\underline{N}^{T}\underline{\vec{R}}^{T}
\end{equation}
and $\underline{A}^{T}$ is the transpose of matrix $\underline{A}$.
%\begin{equation}
%\begin{pmatrix}
%d_{0} \\
%d_{1}
%\end{pmatrix}=
%\underline{R}\underline{N}\underline{I}\underline{M}
%\begin{pmatrix}
%a_{0} \\ 
%a_{1}e^{-i\omega \tau }
%\end{pmatrix}
%\end{equation}.

%\begin{equation}
%\begin{pmatrix}
%d_{0} \\ 
%d_{1}
%\end{pmatrix}
%=
%\begin{pmatrix}
%R_{0} & R_{1} \\ 
%R_{1} & R_{0}
%\end{pmatrix}
%\begin{pmatrix}
%N_{00} & N_{10} \\ 
%N_{01} & N_{11}
%\end{pmatrix}
%\begin{pmatrix}
%I_{0} & I_{1} \\ 
%I_{1} & I_{0}
%\end{pmatrix}
%\begin{pmatrix}
%M_{00} & M_{10} \\ 
%M_{01} & M_{11}
%\end{pmatrix}
%\begin{pmatrix}
%a_{0} \\ 
%a_{1}e^{-i\omega \tau }
%\end{pmatrix}
%\end{equation}
The number of photons
emitted in a given harmonic is proportional to
$\vec{d}_{0}\cdot\vec{d}^{*}_{0}+\vec{d}_{1}\cdot{\vec{d}^{*}_{1}}$. 
The high harmonic intensity is a sum over all Raman
active modes $i$: 
\begin{equation}
P(\tau )=P_{0}+\Sigma _{i}P_{1}^{(i)}\cos \left(\omega
_{i}\tau+\delta _{i}\right)
\label{eq:intensity}
\end{equation}.  
The static $P_{0}$ primarily results from terms of the form
$a_{0}^{*}a_{0}$, while $P_{1}$ results from terms of the form
$a_{0}a_{1}^{*}e^{i \omega \tau}$ and $a_{1}a_{0}^{*}e^{-i \omega
  \tau}$.  Defining
$\underline{W}=\underline{M}^{\dag}\underline{I}^{\dag}
\underline{N}^{\dag} 
\underline{\vec{R}}^{\dag}\cdot \underline{\vec{R}NIM}$,
$P_{0}=a_{0}^{*}W_{00}a_{0}^{*}$ and $P_{1}\cos \left( \omega
t+\delta \right)=\frac{1}{2}(a_{1}^{*}e^{i
  \omega \tau}W_{10}a_{0}+\text{c.c.})$.  Since $I_{1}$ and
$R_{1}$ are small relative to $I_{0}$ and $R_{0}$, only their
first-order terms are kept.

At this level of approximation, calculating
$\hat{I}$ and $\hat{\vec{R}}$ as 
functions of the nuclear coordinates and substituting the expectation
values for $Q_{i}$ at ionization and recombination would give
identical results.  Nevertheless, tracking the
quantum mechanical pathways in this manner is informative, because it
allows the prediction of other observables less amenable to a
``classical nuclear motion'' analysis, like the relative populations
in $\ket{0}$ or $\ket{1}$ after recombination.

%The preceding development gives our framework for analyzing the modulations
%of HHG intensity that occur in the pump-probe experiment of Wagner et al.
%This framework gives identical results to this order in the calculation to
%the HHG total intensity at any given photon energy that would be calculated
%by first calculating the HHG intensity as a function of the nuclear
%coordinates, followed by insertion of the time-dependent expectation values $%
%\left\langle Q_{i}(\tau )\right\rangle .$ \ Nevertheless, we view it as
%potentially informative to break down the quantum mechanical pathways in the
%manner described above, because it allows the prediction of other
%observables less amenable to the ``classical nuclear motion'' analysis, such
%as the intensity of that subset of the harmonic photons that leave the final
%SF$_{6}$ in a specific vibrational state.\ \ Next we carry out a
%quantitative evaluation of the various ``beamsplitter amplitudes'', relevant
%to the Wagner et al. experiment.

This framework is applied to real molecules, using an
improved version of the 3 step model.  For the ionization
step, a simple
one-dimensional WKB tunneling picture describes an electron tunneling
only in directions parallel to the laser electric field.  This is
motivated by the ``initial value representation''
\cite{miller2001,nakamura2005}. 
In the classically forbidden region under the barrier formed by the
molecular potential and the laser field, the tunneling
wavefunction equals the value of the unperturbed molecular HOMO at
the inner turning point times a declining WKB exponential.
With the direction of the electric field
as $-\hat{z}$, the wavefunction at the outer and inner turning
points are related by 
\begin{equation}
\begin{split}
&\psi_{t} (x,y,z_{\text{tp2}},t) =\psi_{\text{HOMO}} (x,y,z_{\text{tp1}},t)\times  \\
&|C_{1}/C_{2}|^{1/6}\text{Bi}(0)/\text{Ai}(0)
\exp[-\int_{z_{\text{tp1}}}^{z_{\text{tp2}}}dz k(z)],
\end{split}
\end{equation}
where $k(z)=\sqrt{2(V_{\text{mol}}+V_{\text{laser}}-E)}$, Ai and Bi
are Airy functions,
$z_{\text{tp1}}$ and $z_{\text{tp2}}$ are the inner and outer turning
points, $C_{1}$ and $C_{2}$ are the z components of the slopes of $2
V(\vec{r},t)$ at the two turning points, and the path integral is
calculated along the $z$ direction parallel to the applied electric
field.

%The second step of the HHG process refers to the time interval between
%ionization and recombination. During this time the evolution of the
%electron's wavefunction is relatively simple and can be treated
%semiclassically. 

After tunneling, the free electron wavefunction's evolution is
relatively simple until it rescatters from the parent ion.
$\psi _{c}(r,\Omega ,t)$, the continuum
wavefunction at the 
instant just prior to the electron rescattering from the molecular
ion, is found
using Gutzwiller's semiclassical propagator \cite{gutzwiller}: 
\begin{equation}
\begin{split}
K(\vec{r},t;\vec{r}_{0},t_{0})& =(2\pi i)^{-3/2}\sqrt{C(\vec{r},t;\vec{r}%
_{0},t_{0})}\times  \\
& \exp [iS(\vec{r},t;\vec{r}_{0},t_{0})-i\phi ]
\end{split}
.
\end{equation}
Here $S(\vec{r},t;\vec{r}_{0},t_{0})$ is the action integral $S=\int L(q,%
\dot{q},t)dt$ calculated for the classical trajectory starting at $(\vec{r}%
,t)$ and ending at $(\vec{r}_{0},t_{0})$ and $C(\vec{r},t;\vec{r}%
_{0},t_{0})=|-\frac{\partial ^{2}S}{\partial r_{0,A}\partial r_{B}}|$ is the
density of trajectories for given initial and final points. $\phi $ is a
phase factor equal to $\frac{\pi }{2}$ times the number of conjugate points
crossed by the trajectory.  The semiclassical continuum
wavefunction is
\begin{equation}
\psi_{c}(\vec{r},t)=\int d^3 \vec{r}_{0}\int d t_{0}
K(\vec{r},t;\vec{r}_{0},t_{0}) \psi_{t}(\vec{r}_{0},t_{0})
\label{eq:psic}
\end{equation}

When the electron recollides with the parent ion, its
wavefunction is distorted strongly by the molecular potential and by
exchange effects with the other electrons, which can
dramatically change amplitudes for recombination with respect to the 
plane wave approximation.
Techniques described in Refs
\cite{tonzani2005,tonzani2006a,tonzani2006b} determine a complete set of
stationary field-free electron-molecule scattering states. Beyond the
range of the molecular potential, the scattering states are given in terms of
incoming and outgoing Coulomb radial functions $f^{\pm}_{El}(r)$ and
the scattering S-matrix as
\begin{equation}
\begin{split}
\psi _{E,lm}(\vec{r})=& \frac{1}{i\sqrt{2}}f_{El}^{-}(r)Y_{l,m}(\theta
,\phi )- \\
& \frac{1}{i\sqrt{2}}\sum_{l^{\prime },m^{\prime }}f_{El^{\prime
}}^{+}(r)Y_{l^{\prime },m^{\prime }}(\theta ,\phi )S_{l^{\prime },m^{\prime
};l,m}(E).
\end{split}
\label{eq:scatteringstates}
\end{equation}
During the electron-molecule scattering, when
recombination occurs, the electron wavepacket is expanded in terms of
these scattering states as   
$\psi _{\text{S}}=\int dE\sum_{l,m}A_{l,m}(E)\psi _{E,lm}e^{-iEt}$,
where 
\begin{equation}
A_{l,m}(E)=e^{iEt}\int d^3 \vec{r}\psi _{E,lm}^{\ast
}(\vec{r})\psi _{c}(\vec{r},t).  \label{eq:alm}
\end{equation}

For a chosen time of projection onto the scattering states,
Eqs. (\ref{eq:psic}) and (\ref{eq:alm}) together define a
seven-dimensional integral over initial and final positions and
initial times.  However, the integrand oscillates rapidly almost
everywhere, causing cancellations. 
Stationary phase techniques identify the region where the integrand
oscillates slowly, which permits evaluation.

%We first expand the semiclassical action $R$
%to second order around the starting and ending points of some
%classical trajectory, and about the starting time.
%$R=R_{0}+R_{x_{A}}\delta x_{A}+\frac{1}{2}R_{x_{A},x_{B}}\delta x_{A}
%  \delta x_{B}+R_{x_{0,A}}\delta x_{0,A}+\frac{1}{2}R_{x_{0,A},x_{0,B}}
%  \delta x_{0,A} \delta x_{0,B} + R_{t_{0}} \delta t_{0}+
%\frac{1}{2}R_{t_{0},t_{0}}(\delta t_{0})^{2}$.  The identities
%$\frac{\partial R}{\partial r_{0,A}}=-p_{0,a}$,$\frac{\partial
%  R}{\partial r_{A}}=p_{a}$ and $\frac{\partial
%  R}{\partial{t_{0}}}=\hat{H}(t_{0})$ (\cite{gutzwiller}, see also
%  \cite{goldstein} section 8.6) are very helpful in this type of
%  expansion.  
The semiclassical action $S$ is expanded to second order around the
starting and ending points of some classical trajectory, and about the
starting time.
Near the starting position and time, $\psi_t(\vec{r}_{0}+\delta
  \vec{r}_{0},t_{0}+\delta
  t_{0})=\psi_t(\vec{r}_{0},t_{0})exp[i(\vec{k}_{0} \delta
\vec{r}_{0}-iE_{\text{HOMO}} \delta t_{0})]$, while
$f^{\pm}_{E,l}(r+\delta r)=f_{E,l}(r)exp[\mp k_{E,l}(r) \delta r]$,
where $k_{E,l}(r)=\sqrt{2(E-V_{l}(r))}$, near the final position.
Angular derivatives of the spherical
harmonics are neglected.   

%Clearly, the phase oscillations of the integrand occur most slowly
%when the linear terms in the $R$ expansion for some trajectory are 
%canceled by the phase evolution of the initial or final states.  These
%stationary phase trajectories are not the only
%trajectories of interest in the problem; rather they are the
%trajectories that best describe the phase in the slowly-oscillating
%region of interest.

The integrand oscillates most slowly when its complex phase is
nearly constant.  This happens when the linear terms in the $S$
expansion are canceled by the linear terms in the complex phases of
the initial and final states.  For such a trajectory, the
contributions of nearby trajectories with nearly the same
$\vec{x}_{0},t_{0},\vec{x},t$ to the integral will tend to add constructively.
These ``stationary phase'' trajectories are not the only trajectories
of interest in the problem, but expanding S about their beginning and
ending points describes the phase in the slowly-varying region of interest.

These stationary phase conditions are met by a trajectory that begins
``downstream'' of the molecule in the direction of the electric field
with zero momentum, moves only radially and parallel to the electric
field until it reencounters the parent molecule at the scattering
state energy.  Only the incoming-wave part of the scattering states
gives a nonvanishing contribution to the expansion coefficients.
Performing the resulting gaussian integrals about the 
initial and final points, and about the initial time, yields expansion
coefficients
\begin{equation}
\begin{split}
A_{l,m}(E)=\frac{4 \pi^2 }{i \sqrt{2}} \sqrt{\left |\frac{\partial r_{A}}{\partial
    p_{0,B}} \right |} \left(\frac{\partial H}{\partial t_{0}} \right)^{-\frac{1}{2}}
    f_{El}^{-*}(r)Y^{*}_{l,m}(\hat{z}) \\
    e^{i[S(\vec{r},t,\vec{r}_{0},t_{0})-\phi]} \psi_{t}(\vec{r}_{0})
    e^{[-iE_{\text{HOMO}} t_{0}]}
\end{split}
\end{equation}
for a stationary phase trajectory that starts at $(z_{0},t_{0})$
and ends at 
$(z,t)$, where $\hat{z}$ is the direction of the electric field
at the time of ionization.

The recombination amplitude is now
$\vec{D}(E)=\sum_{l,m}A_{l,m}(E)\vec{d}_{l,m}(E)$, where
$\vec{d}_{E,lm}=\bra{\psi_{g}}\vec{x}\ket{\psi_{E,lm}}$ is the
recombination amplitude calculated for each individual
scattering state.  This connects to the quantum paths framework when 
$I=\psi_t(\vec{r}_0,t_{0})$, and $\vec{R}$=$\vec{D}(E)$ calculated
for $I=1$.  Both quantities are calculated at the equilibrium
geometry, then for a geometry distorted by 0.1 bohr in the normal mode
coordinate to find $\hat{I}$,$\hat{\vec{R}}$ and their
derivitives.  This involves recalculating the scattering states and
recombination dipoles for the distorted geometry.

%For each scattering state $\psi_{E,lm}$, we calculate a recombination
%dipole amplitude vector
%$\vec{d}_{E,lm}=\bra{\psi_{g}}\vec{x}\ket{\psi_{E,lm}}$.  Then the
%recombination amplitude vector for the scattering wavefunction is 
%$\vec{D}(E)=\sum_{l,m}A_{l,m}(E)\vec{d}_{l,m}(E)$.  
%We relate this to
%our quantum paths model of the vibrational wavepacket by setting
%$I=\psi_t(\vec{r}_0,t_{0})$, and $\vec{R}$
%equal to $\vec{D}(E)$ calculated for $I=1$.  We calculate both
%quantities at the equilibrium geometry to get $I_{0},R_{0}$.  We find
%the derivatives of these quantities by distorting the molecule by 0.1
%bohr in the normal mode coordinate and calculating $I$ and $R$ for the
%new geometry.  This involves recalculating the scattering states and
%recombination dipoles for the distorted geometry.

\begin{figure}
\begin{center}
\includegraphics[width=3.3in]{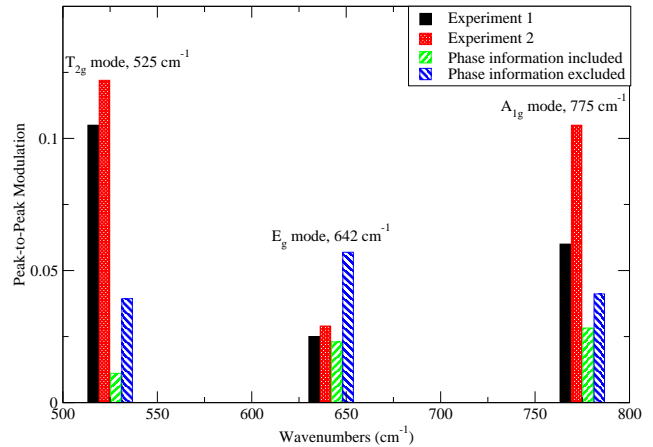}
\end{center}
\caption{(Color online) Peak-to-peak modulation of the high harmonic signal
vs. wavenumber, comparing theory to the two experimental runs for
which data is available.  Modulations corresponding to the same
frequency have been placed side-to-side for purpose of comparison.}
\label{fig:theoryexptcomparison}
\end{figure}

The modulation of the 39th harmonic, which the JILA experiment
considered in detail, is calculated for comparison with experiment.
This harmonic is close
to the measured cutoff, so it can only be produced by a laser
half-cycle where the maximum of the electric field falls close to the
maximum of the Gaussian envelope.
Accordingly, only the modulation for a single half-cycle where these two
maxima coincide was used.  In the expression for the harmonic intensity at some
harmonic order, Eq. (\ref{eq:intensity}), all the quantities depend on
molecular orientation.  
A rotational average of $P_{0}$ and $P_{1}$ were calculated, since the
JILA experiment was performed on a gas jet of molecules which had no
preferred orientation.  Only polarizations perpendicular to the
propagating beam were included in these averages.  Although each of
the T$_{2g}$ and 
$E_{g}$ modes modulates the harmonic intensity strongly at particular
orientations, the phase offset $\delta$ in $P_{1}cos(\omega \tau
+\delta)$ changes with orientation, cancelling some of the observed
oscillation.  The more symmetric A$_{g}$ mode experiences less
cancellation because the initial Raman pulse stimulates it equally for
all molecular orientations.   The trends in the peak intensities of
different normal modes
are, however, similar with and without the phase.  Fig.
\ref{fig:theoryexptcomparison} compares the spherically
averaged peak-to-peak modulation with phase information included, with
phase information excluded by setting $\delta=0$ for all orientations,
and the experimentally measured modulations for each mode for the two
runs for which all three mode modulations could be distinguished from
the background \cite{wagnerpersonal}. It should be noted
that although the relative peak intensities are not in
agreement with experiment, the intensities for the different peaks are all the
same order of magnitude, whereas in the Raman spectrum the $A_g$ mode is 20
times more intense than the others. \cite{wagner2006}

Since the scattering wavefunction is expanded in field-free scattering
states, the calculated modulation varies slightly depending on 
the time at which the semiclassical wavefunction is projected onto the
scattering states. Additional uncertainty may arise because of the various
approximations made in the treatment of molecular scattering, detailed in 
\cite{tonzani2005,tonzani2006a,tonzani2006b}.
Fig. \ref{fig:theoryexptcomparison} shows modulations calculated
when this projection is made at $\omega t=3.9$ radians, when the
short electron trajectories return to the vicinity of the
parent molecule with the correct energy to yield 39th harmonic photons
upon recombination. 
%The experimental
%bars show the measured modulation for the two experimental runs where
%the vibrational modulation could be distinguished from the background
%at all three frequencies\cite{wagnerpersonal}. 

%The internal degrees of freedom present in molecules thus allow for
%phenomena in HHG that have no analogues in atomic systems.  We have
%demonstrated that the HHG signal can be modulated interferometrically
%by a molecule's vibrational state.  Thus
%HHG could serve both as an interferometric probe of chemical
%dynamics (similar to the way in which rearrangement dynamics is investigated in
%Ref. \cite{baker2006}) and as a complementary tool to traditional spectroscopic
%techniques, offering great potential for
%future investigations.

%In the present work we have combined for the first time a rigorous treatment of
%electronic scattering with a semiclassical description of the laser field
%effect on the scattering states, resulting in a flexible and robust
%implementation that has been used to treat a complex molecule, such as
%SF$_6$, with many internal degrees of freedom to a level of sophistication that
%is unprecedented in this area of research for such a large system. The results show signal modulation
%due to vibrational motion, with the correct order of magnitude, the A$_g$ mode
%peak intensity is comparable to the other two modes (unlike in the Raman
%spectra where it is 20 times more intense)  although the
%relative peak intensities for the different modes are not in agreement with
%experiment.

In the present work, we have combined for the first time a rigorous
treatment of electron-molecule scattering with a semiclassical
description of electronic propagation, resulting in a flexible and
robust implementation that has been used to treat a complex molecule
with many internal degrees of freedom to a level of sophistication
that is unprecedented in this area of research for such a large
system.  We have shown how the internal degrees of freedom allow for
phenomena which have no analogue in atomic systems, demonstrating that
the HHG signal can be modulated interferometrically by a molecule's
vibrational state.  Thus HHG could serve both as an interferometric
probe of chemical dynamics (similar to the way in which rearrangement
dynamics are investigated in Ref. \cite{baker2006}) and as a
complementary tool to traditional spectroscopic techniques, offering
great potential for future investigations.

{\em Acknowledgments:} The authors would like to thank N. Wagner,
A. W\"uest, M. Murnane, and H. Kapteyn for many stimulating
discussions.  This work was 
supported in part by the Department of Energy, Office of Science, and
in part by the NSF EUV Engineering Research Center.

%%%Figures

\bibliographystyle{apsrev}
%\bibliography{quantumbeats_mk3}

\begin{thebibliography}{15}
\expandafter\ifx\csname natexlab\endcsname\relax\def\natexlab#1{#1}\fi
\expandafter\ifx\csname bibnamefont\endcsname\relax
  \def\bibnamefont#1{#1}\fi
\expandafter\ifx\csname bibfnamefont\endcsname\relax
  \def\bibfnamefont#1{#1}\fi
\expandafter\ifx\csname citenamefont\endcsname\relax
  \def\citenamefont#1{#1}\fi
\expandafter\ifx\csname url\endcsname\relax
  \def\url#1{\texttt{#1}}\fi
\expandafter\ifx\csname urlprefix\endcsname\relax\def\urlprefix{URL }\fi
\providecommand{\bibinfo}[2]{#2}
\providecommand{\eprint}[2][]{\url{#2}}

\bibitem[{\citenamefont{Wagner et~al.}(2006)\citenamefont{Wagner, Wuest,
  Christov, Popmintchev, Zhou, Murnane, and Kapteyn}}]{wagner2006}
\bibinfo{author}{\bibfnamefont{N.~L.} \bibnamefont{Wagner}},
  \bibinfo{author}{\bibfnamefont{A.}~\bibnamefont{Wuest}},
  \bibinfo{author}{\bibfnamefont{I.~P.} \bibnamefont{Christov}},
  \bibinfo{author}{\bibfnamefont{T.}~\bibnamefont{Popmintchev}},
  \bibinfo{author}{\bibfnamefont{X.}~\bibnamefont{Zhou}},
  \bibinfo{author}{\bibfnamefont{M.~M.} \bibnamefont{Murnane}},
  \bibnamefont{and} \bibinfo{author}{\bibfnamefont{H.~C.}
  \bibnamefont{Kapteyn}}, \bibinfo{journal}{Proc. Natl. Acad. Sci. U.S.A.}
  \textbf{\bibinfo{volume}{103}}, \bibinfo{pages}{13279}
  (\bibinfo{year}{2006}).

\bibitem[{\citenamefont{Lewenstein et~al.}(1994)\citenamefont{Lewenstein,
  Balcou, Ivanov, L'Huillier, and Corkum}}]{lewenstein}
\bibinfo{author}{\bibfnamefont{M.}~\bibnamefont{Lewenstein}},
  \bibinfo{author}{\bibfnamefont{P.}~\bibnamefont{Balcou}},
  \bibinfo{author}{\bibfnamefont{M.Y.}~\bibnamefont{Ivanov}},
  \bibinfo{author}{\bibfnamefont{A.}~\bibnamefont{L'Huillier}},
  \bibnamefont{and} \bibinfo{author}{\bibfnamefont{P.B.}~\bibnamefont{Corkum}},
  \bibinfo{journal}{Phys. Rev. A} \textbf{\bibinfo{volume}{49}},
  \bibinfo{pages}{2117} (\bibinfo{year}{1994}).

\bibitem[{\citenamefont{Tonzani and Greene}(2005)}]{tonzani2005}
\bibinfo{author}{\bibfnamefont{S.}~\bibnamefont{Tonzani}} \bibnamefont{and}
  \bibinfo{author}{\bibfnamefont{C.~H.} \bibnamefont{Greene}},
  \bibinfo{journal}{J. Chem. Phys.} \textbf{\bibinfo{volume}{122}},
  \bibinfo{pages}{014111} (\bibinfo{year}{2005}).

\bibitem[{\citenamefont{et~al}(1998)}]{g98}
\bibinfo{author}{\bibfnamefont{M.~J.~F.} \bibnamefont{et~al}},
  \bibinfo{journal}{Gaussian Inc., Pittsburgh, PA}  (\bibinfo{year}{1998}).

\bibitem[{\citenamefont{Dunning}(1989)}]{dunning1989}
\bibinfo{author}{\bibfnamefont{T.}~\bibnamefont{Dunning}}, \bibinfo{journal}{J.
  Chem. Phys} p. \bibinfo{pages}{1007} (\bibinfo{year}{1989}).

\bibitem[{\citenamefont{Estreicher and Estle}(1985)}]{estreicher}
\bibinfo{author}{\bibfnamefont{S.}~\bibnamefont{Estreicher}} \bibnamefont{and}
  \bibinfo{author}{\bibfnamefont{T.~L.} \bibnamefont{Estle}},
  \bibinfo{journal}{Phys. Rev. B} \textbf{\bibinfo{volume}{31}},
  \bibinfo{pages}{5616} (\bibinfo{year}{1985}).

\bibitem[{\citenamefont{Moffitt and Thorson}(1957)}]{moffitt}
\bibinfo{author}{\bibfnamefont{W.}~\bibnamefont{Moffitt}} \bibnamefont{and}
  \bibinfo{author}{\bibfnamefont{W.}~\bibnamefont{Thorson}},
  \bibinfo{journal}{Phys. Rev.} \textbf{\bibinfo{volume}{108}},
  \bibinfo{pages}{1251} (\bibinfo{year}{1957}).

\bibitem[{\citenamefont{Bersuker}(1984)}]{bersuker}
\bibinfo{author}{\bibfnamefont{I.}~\bibnamefont{Bersuker}},
  \emph{\bibinfo{title}{The Jahn-Teller effect and vibronic interactions in
  modern chemistry}} (\bibinfo{publisher}{Plenum Press}, \bibinfo{year}{1984}).

\bibitem[{\citenamefont{Miller}(2001)}]{miller2001}
\bibinfo{author}{\bibfnamefont{W.~H.} \bibnamefont{Miller}},
  \bibinfo{journal}{J. Phys. Chem. A} \textbf{\bibinfo{volume}{105}},
  \bibinfo{pages}{2942} (\bibinfo{year}{2001}).

\bibitem[{\citenamefont{Nakamura}(2005)}]{nakamura2005}
\bibinfo{author}{\bibfnamefont{H.}~\bibnamefont{Nakamura}},
  \bibinfo{journal}{J. Theor. Comp. Chem.} \textbf{\bibinfo{volume}{4}},
  \bibinfo{pages}{127} (\bibinfo{year}{2005}).

\bibitem[{\citenamefont{Gutzwiller}(1990)}]{gutzwiller}
\bibinfo{author}{\bibfnamefont{M.~C.} \bibnamefont{Gutzwiller}},
  \emph{\bibinfo{title}{Chaos in Classical and Quantum Mechanics}}
  (\bibinfo{publisher}{Springer}, \bibinfo{address}{New York},
  \bibinfo{year}{1990}).

\bibitem[{\citenamefont{Tonzani and Greene}(2006)}]{tonzani2006a}
\bibinfo{author}{\bibfnamefont{S.}~\bibnamefont{Tonzani}} \bibnamefont{and}
  \bibinfo{author}{\bibfnamefont{C.~H.} \bibnamefont{Greene}},
  \bibinfo{journal}{J. Chem. Phys.} \textbf{\bibinfo{volume}{124}},
  \bibinfo{pages}{054312} (\bibinfo{year}{2006}).

\bibitem[{\citenamefont{Tonzani}(2007)}]{tonzani2006b}
\bibinfo{author}{\bibfnamefont{S.}~\bibnamefont{Tonzani}},
  \bibinfo{journal}{Comp. Phys. Comm.} \textbf{\bibinfo{volume}{176}},
  \bibinfo{pages}{146} (\bibinfo{year}{2007}).

\bibitem[{wag()}]{wagnerpersonal}
\bibinfo{note}{N. Wagner, private communication.}

\bibitem[{\citenamefont{Baker et~al.}(2006)\citenamefont{Baker, Robinson,
  Haworth, Teng, Smith, Chirila, Lein, Tisch, and Marangos}}]{baker2006}
\bibinfo{author}{\bibfnamefont{S.}~\bibnamefont{Baker}},
  \bibinfo{author}{\bibfnamefont{J.~S.} \bibnamefont{Robinson}},
  \bibinfo{author}{\bibfnamefont{C.~A.} \bibnamefont{Haworth}},
  \bibinfo{author}{\bibfnamefont{H.}~\bibnamefont{Teng}},
  \bibinfo{author}{\bibfnamefont{R.~A.} \bibnamefont{Smith}},
  \bibinfo{author}{\bibfnamefont{C.~C.} \bibnamefont{Chirila}},
  \bibinfo{author}{\bibfnamefont{M.}~\bibnamefont{Lein}},
  \bibinfo{author}{\bibfnamefont{J.~W.~G.} \bibnamefont{Tisch}},
  \bibnamefont{and} \bibinfo{author}{\bibfnamefont{J.~P.}
  \bibnamefont{Marangos}}, \bibinfo{journal}{Science}
  \textbf{\bibinfo{volume}{312}}, \bibinfo{pages}{424} (\bibinfo{year}{2006}).

\end{thebibliography}

\end{document}